\documentclass[pre,twocolumn, superscriptaddress, showkeys, titlepage, aps, 10pt]{revtex4-1}
\pdfoutput=1

\usepackage{amsmath}
\usepackage{amssymb}
\usepackage{graphicx}
\usepackage{color}
\usepackage{hyperref}

\usepackage{color}

\begin{document}

\title{The many facets of community detection in complex networks}
\author{Michael T. Schaub}
\email{mschaub@mit.edu}
\thanks{corresponding author}
\affiliation{Institute for Data, Systems, and Society, Massachusetts Institute of Technology, Cambridge, MA 02139, USA }
\affiliation{ICTEAM, Universit\'e catholique de Louvain,  B-1348 Louvain-la-Neuve, Belgium} 
\affiliation{naXys and Department of Mathematics, University of Namur,  B-5000 Namur, Belgium}
\author{Jean-Charles Delvenne}
\affiliation{ICTEAM, Universit\'e catholique de Louvain,  B-1348 Louvain-la-Neuve, Belgium} 
\affiliation{CORE, Universit\'e catholique de Louvain,  B-1348 Louvain-la-Neuve, Belgium} 
\email{jean-charles.delvenne@uclouvain.be}
\author{Martin Rosvall}
\email{martin.rosvall@umu.se}
\affiliation{Integrated Science Lab, Department of Physics, Ume{\aa} University, SE-901 87 Ume{\aa}, Sweden}
\author{Renaud Lambiotte}
\email{renaud.lambiotte@unamur.be}
\affiliation{naXys and Department of Mathematics, University of Namur,  B-5000 Namur, Belgium}

\begin{abstract}
Community detection, the decomposition of a graph into essential building blocks, has been a core research topic in network science over the past years.
Since a precise notion of what constitutes a community has remained evasive, community detection algorithms have often been compared on benchmark graphs with a particular form of assortative community structure and classified based on the mathematical techniques they employ. 
However, this comparison can be misleading because apparent similarities in their mathematical machinery can disguise different goals and reasons for why we want to employ community detection in the first place.
Here we provide a focused review of these different motivations that underpin community detection. 
This problem-driven classification is useful in applied network science, where it is important to select an appropriate algorithm for the given purpose.
Moreover, highlighting the different facets of community detection also delineates the many lines of research and points out open directions and avenues for future research.
\end{abstract}

\maketitle
\section{Introduction}

Sparked by the work of Newman and Girvan~\cite{Newman2004,Newman2006} on Modularity in Complex Systems, the area of community detection has become one of the main pillars of network science research.
The promise that we can gain a deeper understanding of a system by discerning important structural patterns within a network has spurred a huge number of studies in this area.
However, as has become abundantly clear by now, this problem has no canonical solution.
In fact, even a general definition of what constitutes a community is still lacking.
The reasons for this are not only grounded in the computational difficulties of tackling community detection. 
Furthermore, various research areas view community detection from different perspectives, which the lack of a consistent terminology illustrates:
`network clustering', `graph partitioning', `community', `block' or `module detection' all carry slightly different connotations.
This jargon barrier creates confusion as soon as readers and authors have different preconceptions and intuitive notions are not made explicit.

We argue that community detection should not be considered as a well-defined problem, but rather as an umbrella term with many facets.
These facets emerge from different goals and motivations of what it is about the network that we want to understand or achieve, which lead to different perspectives on how to formulate the problem of community detection.
Therefore, it is critical to be aware of these underlying motivations when selecting and comparing community detection methods.
Thus, rather than an in-depth discussion of the technical details of different algorithmic implementations~\cite{Schaeffer2007,Fortunato2010,Coscia2011,Parthasarathy2011,Newman2012,Malliaros2013,Xie2013,Fortunato2016}, here we focus on the conceptual differences between different perspectives on community detection. 

By providing a problem-driven classification, however, we do \emph{not} argue that the different perspectives are unrelated.
In fact, in some situations, different mathematical problem formulations can lead to similar algorithms and methods, and the different perspectives can offer valuable insights.
For example, for undirected networks, optimizing the objective function Modularity~\cite{Newman2004}, initially proposed from a clustering perspective, can be interpreted both as optimizing a particular stochastic block model~\cite{Newman2016} and a particular diffusion process on the networks~\cite{Delvenne2013},
In other situations, however, such relations are not apparent. 

Neither do we argue that there is a particular perspective that is \emph{a priori} better suited for any given network.
In fact, no method can consistently perform best on all kinds of networks~\cite{Peel2016}.
Community detection is an unsupervised learning task and we cannot know what are the quantities of interest for the analysis.
Instead, to understand how useful a particular method is, we must take into account the context of why the researcher is interested in the communities~\cite{VonLuxburg2012}.

In the following, we unfold different aims underpinning community detection and discuss how the resulting problem perspectives relate to various applications.
We focus on four broad perspectives that have served as motivation for community detection in the literature: (\emph{i})~community detection as minimization of some form of constraint violation; (\emph{ii}) community detection framed as a discretised analogue of data clustering, in which densely knit groups of nodes are to be found; (\emph{iii}) community detection aiming to identify structurally equivalent nodes in a network, leading to notions such as stochastic block models; and (\emph{iv}) community detection looking for simplified descriptions of the dynamical flows occurring on the network, that is, some form of dynamical model reduction (see Figure 1). While this categorization is not unique, we believe that it can help clarifying concepts about community detection and be a guide to using an appropriate method for a particular purpose.

\begin{figure*}[htb]
    \centering
    \includegraphics[]{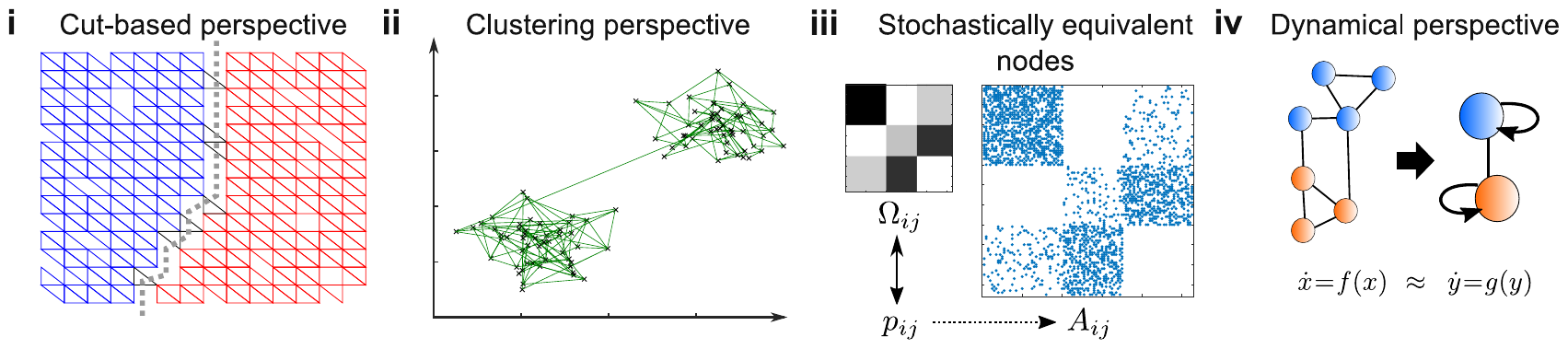}
    \caption{\textbf{Schematic of four different approaches to community detection.}  (\emph{i}) The cut based perspective aims at minimising the number of links between modules, independently of their intrinsic structure. (\emph{ii}) The clustering perspective produces groups of densely connected modules. (\emph{iii}) The stochastic equivalence perspective looks for modules in which nodes are stochastically equivalent, typically as inferred via a generative statistical graph model. (\emph{iv}) The dynamical perspective focuses on the impact of communities on dynamical processes and searches for dynamically relevant coarse grained descriptions.}
    \label{fig:schematics}
\end{figure*}

\section{Minimizing constraint violations: a cut based perspective}\label{sec:II}
One of the earliest graphs partitioning applications was in the area of circuit layout and design~\cite{Alpert1995,Fortunato2010}.
This spurred the development of the now classical Kernighan-Lin algorithm~\cite{Kernighan1970} and the work by Donath and Hoffmann~\cite{Donath1972,Donath1973}, who were among the first to suggest the use of eigenvectors for graph partitioning.
For instance, we might be confronted with a graph which describes the signal flows between different components of a circuit.
To implement the circuit in an efficient way, our goal is now to partition the graph into a fixed number of approximately equally sized groups with a small number of edges between those groups.
The edges that run between the groups are commonly denoted as the cut.
Our aim is thus to minimize this cut while keeping some kind of \textit{balanced groups}, which is an important ingredient in this context.

To make this more precise let us consider one specific variant of this scheme, known as \textit{ratio cut}~\cite{Hagen1992}.
Let us denote the adjacency matrix of an undirected network with $n$ nodes by $A$, where $A_{ij} = 1$ if there is a connection from node $i$ to node $j$, and $A_{ij} =0$ if the nodes are not connected. 
We can now write the problem of optimizing the ratio cut for a bipartition of all vertices $\mathcal V$ into two communities $\mathcal V_1$ and $\mathcal V_2 = \mathcal{V}\backslash\mathcal{V}_1$ as follows~\cite{Hagen1992,VonLuxburg2007}:
\begin{flalign}
    \min_{\mathcal V_i} \text{RatioCut}(\mathcal V_1, \mathcal V_2) := \min_{\mathcal V_i} \sum_i \dfrac{\text{cut}(\mathcal V_i,\mathcal V \backslash \mathcal V_i)}{|\mathcal V_i|},
\end{flalign}
where $\text{cut}(\mathcal{V}_1,\mathcal{V}_2) := \sum_{i\in\mathcal{V}_1, j \in \mathcal{V}_2} (A_{ij} + A_{ji})/2$ is the sum of the (weighted) edges between the two vertex sets $\mathcal V_1, \mathcal V_2$.
Related problem formulations also occur in the context of parallel computations and load scheduling~\cite{Spielman1996,Pothen1997}, where approximately equally sized portions of work are to be sent to different processors, while keeping the dependencies between those tasks minimal.
Further applications include scientific computing~\cite{Spielman1996,Pothen1997}, where partitioning algorithms can be used to divide the coordinate meshes arising in the context of discretizing and solving partial differential equations.
Image segmentation problems may also be phrased in terms of cut-based measures~\cite{Shi2000,VonLuxburg2007}. 

Investigating these types of problems, has led to many important contributions for partitioning graphs, in particular in relation to spectral methods.
The connection of spectral algorithms to cut-based problem formulations arises naturally by considering relaxations of the original, combinatorially hard discrete optimization problems such as \eqref{eq:ratio_cut}, or other related objective functions such as the average or normalized cuts.
This can be best seen when rewriting the above optimzation problem as follows:
\begin{flalign}\label{eq:ratio_cut}
    \min_f & \quad f^TLf \\
    \text{subject to} & \quad f \perp \mathbf{1} \quad \|f\|= \sqrt{n}\\
    \text{where} & \quad f_i:= 
    \begin{cases}
        -\sqrt{|\mathcal V_2|/|\mathcal V_1|} \quad \text{if } i\in \mathcal V_1\\
        \;\;\;\sqrt{|\mathcal V_1|/|\mathcal V_2|} \quad \text{if } i\in \mathcal V_2
    \end{cases}
\end{flalign}
Here the Laplacian matrix of the network has been defined as $L=D-A$, where $D$ is the diagonal degree matrix with $D_{ii} = \sum_{j}A_{ij}$.
We mention here the seminal work of Fiedler~\cite{Fiedler1973,Fiedler1975}, who realized already in the 70s that the second smallest eigenvalue of the Laplacian is associated to the connectivity of the graph, and that the associated eigenvector can thus be used to compute spectral bi-partitions.
Such spectral ideas led to many influential algorithms and methods, see, for example, von Luxburg~\cite{VonLuxburg2007} for a tutorial on spectral algorithms.

In this cut-based problem formulation, there is no specification as to how the found groups in the partition should be connected internally.  While there is the implicit constraint that the groups must not split into groups with an even smaller cut, there is no specification that the found groups of nodes are densely connected internally.
Indeed, the type of graphs considered in the context of cut-based partitions are often of a mesh or grid like form, and for these kind of graphs several guarantees can be given in terms of the quality of the partitions obtained by spectral algorithms~\cite{Spielman1996}.
While such non-dense groupings emerging from `non-clique structures'~\cite{Schaub2012} can also be dynamically relevant (see section \ref{sec:dynamics}), they are likely to be missed when employing a community notion that focuses on finding dense groupings as discussed next.

\section{Maximizing internal density: the clustering perspective}
\label{sec:clustering}
A different motivation for community detection arises when considered in the context of data clustering.
We use the term clustering, which itself has been a synonym for many different things, in the following sense:
For a set of given data points in a possibly high-dimensional space, the goal here is to partition the points into a number of groups, such that points within a group are `close' (or `similar'); and points in different groups are more distant to each other.
To achieve this goal one often constructs a proximity or similarity graph between the points and tries to group nodes together that are closer to each other than they are to the rest of the graph.
This approach results again in a form of community detection problem where the closeness between nodes is described by the presence and weight of the edges between them.

Although minimizing the cut size and maximizing the internal number of links are closely related, there are differences here contrasting with the cut-based perspective outlined above, pertaining to the typical constraints and search space associated with these objective functions.
First, when employing a clustering perspective there is normally no \textit{a priori} information as to the number of groups we are looking for.
Second, we do not necessarily require the groups to be balanced in any way, rather we would like to find an `optimal' split into densely knit groups irrespective of their relative sizes.

Unsurprisingly, finding an optimal clustering is again a computationally hard problem.
Further, as Kleinberg has shown~\cite{Kleinberg2003}, there are no clustering algorithms satisfying a certain set of intuitive properties we might require from a clustering algorithm in continuous spaces; and similar problems also arise in the discrete setting for clustering of graphs~\cite{Browet2016}.

Nevertheless, there exists a large number of methods that follow a clustering like paradigm and separate the nodes of a graph into cohesive groups of nodes, often by optimizing a quality function.
An important clustering metric in this context is the so called conductance~\cite{Kannan2004,Andersen2006,Spielman2013,Kloster2014}.
Optimizing the (global) conductance has been initially introduced as a way to produce a global bi-partition similarly to the 2-way ratio-cut.
However, more lately this quantity has been successfully employed as a local quality function to find localized clusters around one or more seed nodes.
Given a set of nodes $\mathcal V_k\subset \mathcal V$, a potential community, its local conductance can be written as:
\begin{equation}
    \phi(\mathcal V_k) := \frac{\sum_{i\in \mathcal V_k, j \notin \mathcal V_k}A_{ij}}{\min\{\text{vol}(\mathcal V_k),\text{vol}(\mathcal V \backslash \mathcal V_k )\}},
\end{equation}
where $\text{vol}(\mathcal V_k) := \sum_{i\in \mathcal V_k}\sum_j A_{ij}$ is the total degree of the nodes in set $\mathcal V_k$, commonly termed its volume in analogy with (continuous) geometric objects.
Interestingly, it has been shown that in specific contexts the conductance can be a good predictor of some latent group structures in several real-world applications~\cite{Yang2015}\footnote{Let us emphasize here again that this fact does not mean that conductance is a more meaningful algorithm in any way, or able to reveal some generic `ground truth'; see~\cite{Peel2016} for an extensive discussion on the relation of meta-data and structure}.

Moreover, a local perspective on community detection has two appealing  properties: 
First, the definition of a cluster does not depend on the global graph structure but only on the relative local density.
Second, only a portion of a graph needs to be accessed, which is advantageous if there are computational constraints (very large graphs), or we are only interested in a particular subsystem.
In such cases, we would like to avoid having to apply a method to the whole graph to find, for example, the cluster containing a particular node in the graph.

The Newman-Girvan Modularity~\cite{Newman2004,Newman2006} is arguably one of the most common clustering measures used in the literature and was originally proposed from the clustering perspective discussed here.
It is a global quality function and aims to find the community structure of the network as a whole.
Given a partition $\mathcal P =\{\mathcal V_1, \ldots, \mathcal V_k\}$ of a graph into $k$ groups, the Modularity of $\mathcal P$ can be written as:
\begin{equation}
    Q(\mathcal P) := \dfrac{1}{2m}\sum_{q =1}^k\sum_{i,j \in \mathcal V_q} \left[A_{ij} - \dfrac{d_id_j}{2m}\right],
\end{equation}
where $d_i = \sum_j A_{ij}$ is the degree of node $i$ and ${2m = \sum_i d_i}$ is the total weight of all edges in the graph.
By optimizing the Modularity measure over the space of all partitions, one aims to identify groups of nodes that are more densely connected to each other than one would expect according to a statistical null model of the graph.
This statistical null model is commonly chosen to be the configuration model with preserved degree sequence.

However, a by-product of this choice of a global null-model is the tendency of Modularity to balance the size of the modules in terms of their total connectivity.
While different variants of Modularity aim to account for this effect~\cite{Fortunato2010}, this makes Modularity also interpretable as a trade-off between a cut-based measures and an entropy~\cite{Delvenne2013}.
In fact, Modularity can be seen as a proxy for all the perspectives discussed in this article.
The optimization of Modularity is usually performed by means of spectral or greedy algorithms~\cite{Fortunato2010,Newman2006a,Blondel2008}.
While there are problems with this approach, such as its resolution limit~\cite{Fortunato2007} and other spurious effects~\cite{Fortunato2007,Good2010,Guimera2004,Lancichinetti2011a}, the general idea has triggered development of a plethora of algorithms that follow a similar strategy~\cite{Fortunato2010}.
Several works have addressed some of these shortcomings, for instance by incorporating a resolution parameter, or explicitly accounting for the density inside each group~\cite{Chen2014,Chen2015}.

By grouping \textit{similar} nodes that link to similar nodes into communities, we constrain ourselves to finding \textit{assortative} group structure~\cite{Fortunato2016}.
Stated differently, if we ordered the nodes in the network according to the underlying group structure, the adjacency matrix would be close to block diagonal.
While we may also have hierarchical clusters with clusters of clusters etc., such an assortative structural organisation might be too restrictive if we want to analyse, for example, social networks or capture the organisation of bipartite networks.
If we aim to define groups based on more general connectivity patterns, this leads naturally to notions such as the stochastic equivalence, which we will consider in the next section.

\section{Nodes with similar structural roles and stochastic block models}
Within social network analysis, a common goal is to identify nodes within a network that serve a similar structural role in terms of their connectivity profile.
Accordingly, nodes are similar if they share the same kind of connection patterns to other nodes.
This idea is captured in notions such as \textit{regular equivalence}, which states that nodes are regularly equivalent if they are equally related to equivalent others~\cite{Everett1994,Hanneman2005}.
A relaxation of this idea is \emph{stochastic equivalence}~\cite{Holland1983}, which means that nodes are equivalent if they connect to equivalent nodes with equal probability.

One of the most popular techniques to model and detect such kind of relationship in network data is the use of stochastic block models (SBMs)~\cite{Holland1983,Nowicki2001} and associated inference techniques.
These models have their roots in the social networks literature~\cite{Holland1983,Anderson1992}, and provide a flexible framework for modelling block structures within a network.
When considering block models, we are interested in identifying node groups such that nodes within a community connect to nodes in other communities in an `equivalent way'~\cite{Fortunato2016}.

Consider a network composed of $n$ nodes divided into $k$ classes.
The standard SBM is defined by the set of node class labels and the affinity matrix $\Omega$.
More precisely, the link probability between two nodes $i, j$ belonging to class $c_i$ and $c_j$ is given by: 
$$p_{ij} := \mathbb P(A_{ij})= \Omega_{c_ic_j}.$$
Under an SBM, nodes within the same class thus have exactly the same probabilities to connect to nodes of another class. This is the mathematical formulation of having stochastically equivalent nodes within each class.
Finding the latent groups of nodes in a network now amounts to inferring the model parameters that provide the best fit for the observed network. That is, find the SBM with the highest likelihood.

The standard SBM assumes that the expected degree of each node is a Poisson binomial random variable (a Binomial random variable with possibly non-identical success probabilities in each trial).
Because inferring the most likely SBM typically results in grouping nodes based on their degree in empirical networks with broad degree distributions, it can be advantageous to include a degree correction into the model. 
In the degree corrected SBM~\cite{Karrer2011}, the probability $p_{ij}$ for a link to appear between two nodes $i, j$ depends both on their class labels $c_i, c_j$ and their respective degree parameters $d_i, d_j$ (each entry $A_{ij}$ might be a Bernoulli or a Poisson random variable such as in~\cite{Karrer2011}):
$$p_{ij} \sim d_i d_j \Omega_{c_ic_j}.$$ 
Thus, while edges in real-world networks tend to be correlated from effects such as triadic closure~\cite{Fortunato2010}, by construction edges are conditionally independent random variables in SBMs. 
Moreover, most common SBMs are defined for unweighted networks or networks with integer weights by modelling the network as a multi-graph.
Though there are generalizations~\cite{Aicher2014,peixoto2015inferring}, this is still an area comparably less studied.

In contrast to the notions of community considered above, with stochastic equivalence we are no longer interested in maximising some internal density or minimising a cut. 
To see this, consider a bipartite graph that from a cut- or density-based perspective contains no communities (one may even see bipartite structure as `anti-communities').
From the stochastic equivalence perspective, however, we would say that this graph contains two groups because nodes in each set only connect to nodes in the other set.

When adopting an SBM to detect such structural organisation of the links, we explicitly adopt a statistical model for the networks.
The network is essentially an instance of an ensemble of possible networks generated from such a model~\footnote{This ensemble assumption is also reflected in the Modularity formalism, where the observed network is compared to a null model.}.
This model based approach comes with several advantages:
First, by defining the model we effectively declare what is signal and what is noise in the data under the SBM.
We can thus provide a statistical assessment of the observed data with, for example, $p$-values under the SBM.
In other words, we can identify patterns that cannot be reasonably explained from density fluctuations of edges inherent to any realisation of this model.
Second, we are, for example, able to generate new networks from our model with a similar group structure, or predict missing edges and impute data.
Third, we can make strong statements about the detectability of groups within a graph.
For example, precise criteria specify when any algorithm can recover the planted group structure for a graph created by an SBM~\cite{Decelle2011a,Mossel2013}.
By fitting an SBM to an observed adjacency matrix it is possible to recover such a planted group structure down to its theoretical limit~\cite{Mossel2013,Massoulie2014}.
While these criteria apply to networks generated with SBMs and not real networks in general, in which case we do not know what kind of process created the network~\cite{Peel2016}, it is nevertheless a remarkable result since it highlights that there are networks with undetectable block patterns.

Many benchmark graphs proposed in the literature, such as the commonly used LFR benchmarks~\cite{Lancichinetti2008}, can be seen as specific types of SBMs.
Results on these benchmarks graphs should therefore be interpreted with the SBM perspective in mind, especially with respect to the detectability limit.
Finally, this model based approach also offers ways to estimate the number of communities from the data by some form of model selection, including hypothesis testing~\cite{Bickel2016}, spectral techniques~\cite{Krzakala2013,Saade2014}, the minimum description length principle~\cite{Peixoto2013}, or Bayesian inference~\cite{yan2016bayesian}. 

\section{Communities as dynamical building blocks}
\label{sec:dynamics}
Let us now consider a fourth alternative motivation for community detection, focusing on the processes that take place on the network. 
All notions of community outlined above are effectively structural in the sense that they are mainly concerned with the composition of the graph itself or its representation as an adjacency matrix, respectively.
However, in many cases one of the main goals of applying tools from network science is to understand the \textit{behavior} of a system.
While the topology of a system puts constraints on the dynamics that can take place on the network, the network topology alone cannot explain the system behavior.
Whence, instead of finding a coarse grained description of the adjacency matrix, we might be interested in finding a coarse grained description of the dynamics acting on top of the network.

Take air traffic as an example. 
An airline network, with weighted links connecting cities according to the number of flights between them, can offer some interesting insights about air traffic.
For instance, in the US air traffic network, Las Vegas and Atlanta form two major hubs. 
However, if we instead focus on the passenger flows based on actual itineraries, the two cities show very different behavior: Las Vegas is a tourist destination and typically the final destination of itineraries, whereas Atlanta is a transfer hub onto other final destinations~\cite{Rosvall2014,peixoto2015modeling}. 
Thus, these airports play dynamically quite different roles in the network.
Focusing on interconnection patterns alone can thus give an incomplete picture if we are interested in the dynamical behavior of a system, for which additional dynamical information should be taken into account.
Conversely, a concentration of edges with high impact on the dynamics may just arise from a statistical fluctuation, if the network is seen as a realization of a particular random graph model.
In this way, structural and dynamical approaches can offer complementing information. 

Flow-based community detection approaches focus on specifying the modular dynamics on a given, fixed network.
Consequently, depending on the dynamics of interest, the modular building blocks may  look different. 
In general, however, they are blocks of nodes with different identities that trap the flow or channel it in specific directions. 
That is, they form reduced models of the dynamics where blocks of nodes are aggregated to single meta nodes with similar \textit{dynamical function} with respect to the rest of the network. 
In this view, the goal of community detection is to find effective coarse-grained system descriptions of how the dynamics take place on the network structure.

This dynamical take on community detection has primarily focused on modelling the dynamics with Markovian diffusion processes~\cite{Rosvall2008,Delvenne2010,Lambiotte2014}, though work of topological scales and synchronization share the same common ground~\cite{Arenas2006}.
Interestingly, for a simple diffusion dynamics such as a random walk on an undirected network, which is essentially determined by the spectral properties of the network Laplacian, this perspective is tightly connected with the clustering perspective discussed in section \ref{sec:clustering}.
This is because the presence of densely knit groups within the network can introduce a time-scale separation in the diffusion dynamics:
A random walker traversing the network will initially be trapped for a significant time inside a community corresponding to the fast time-scale, before it can escape and explore the larger network corresponding to a slower time-scale.
However, already for directed networks this connection between link density and dynamical behavior breaks down, even for a simple diffusion process~\cite{Rosvall2008,Lambiotte2014,Schaub2012}.
This apparent relationship breaks down completely when focusing on longer pathways, possibly with memory effects in the dynamics~\cite{Rosvall2014,Salnikov2016}.

A dynamical perspective is useful especially in applications in which the network itself is well defined, but the emergent dynamics are hard to grasp.
For instance, consider the nervous system of the roundworm \textit{C. elegans} for which there exists a distinct network.
A basic generative network model, such as a Barabasi-Albert graph or an SBM, might be too simple to capture the complex architecture of the network, and sampling alternative networks from such a model will not create valid alternative roundworm connectomes. 
Indeed, some more complicated network generative models have been proposed to model the structure of the network~\cite{Nicosia2013}, and may be used to assess the significance of individual patterns compared to the background of the assumed model.
However, if instead we are interested in assessing the dynamical implications of the evolutionary conserved network structure, it may be fruitful to engineer differences in the actual network and investigate how they affect the dynamical flows in the system.
For instance, one can replicate experimental node ablations \textit{in silico} and assess their dynamical impact~\cite{Bacik2016}.

In the dynamical perspective we are typically interested in how short term dynamics are integrated into long term behavior of the system and seek a coarse grained description of the dynamics occurring on a given network.
The network itself represents the true structure, save for empirical imperfections. 
This dynamical viewpoint is not tied to a particular method: for instance, it is possible to formulate generative statistical models for empirically observed pathways~\cite{peixoto2015modeling}\footnote{Note, however, that whereas the generative approach in~\cite{peixoto2015modeling} tries to explicitly model the underlying state space of the trajectories, we may simply be interested in effectively compressing the long term behavior of the system, which is a somewhat different goal. See, for instance, the discussion in Ref.~\cite{peixoto2015modeling,persson2016maps}}.
Compared to some the previous perspectives, the dynamical viewpoint has received somewhat less attention and has been confined mainly to diffusion dynamics.
A key challenge is to extend this perspective to other types of dynamics and link it more formally to approaches of model order reduction considered in control theory.
In light of the recently growing interest in control of complex systems, this could help us better understanding complex systems.

\section{Discussion}
Community detection can be viewed through a range of different lenses. 
Rather than looking at community detection as a generic tool that is supposed to work in a generic context, considering the application in mind is important when choosing between or comparing different methods.
Each of the perspectives outlined above has its own particularities, which may or may not be suitable for the problem of interest.

We emphasize the different perspectives in the following example.
Given a real-world graph generated by a possibly complex random assignment of edges, we assume that we are interested in some particular dynamics taking place on this graph such as epidemic spreading.
We also assume that the graph is structured such that the dynamics exhibit a time-scale separation.
If, for instance, we want to coarse grain an epidemic and identify critical links that should be controlled to confine the epidemics, then it does not matter whether or not random fluctuations generated the modules that induces the time-scale separation. In any case, these modules will be relevant for the dynamics.

Assume now that the same graph encodes interdependency of tasks in a load scheduling problem.
In such a circumstance, a cut-based approach will find a relevant community structure, in that it allows an optimally balanced assignment of tasks to processors that minimises communication between processors. 
These communities may be very different from the ones attached to the epidemic spreading.
In these two cases, we considered a single realisation of the network, and the goal was to extract useful information about its structure, independently of the possible mechanisms that generated it. 

Let us now consider the same network from a stochastic equivalence perspective, and assume for simplicity that the graph is a particular realization of an Erd\H{o}s-R\'enyi graph.
In this case, an approach based on the SBM is expected to declare that there is no significant pattern to be found here at all, as the encountered structural variations can already be explained by random fluctuations rather than by hidden class labels. 
Thus, communities in the SBM picture are defined via the latent variables within the statistical model of the network structure, and not via their impact on the behavior of the system.
In this way, different motivations for community detection can find different answers even for the very same network. 

In addition to the differences \textit{between} these perspectives, there are also variations \textit{within} each perspective.
For instance, distinct plausible generative models such as the standard SBM or the degree corrected SBM will for a given graph lead to different inferred community structure.
Similar variations exist in the dynamical paradigm as well: distinct natural assumptions for the dynamics, such as dynamics with memory or not, uniform across nodes or edges, etc., applied to a given graph will lead to different partitions.
Also different balancing criteria, see section \ref{sec:II}, or different concepts of high internal density, see section \ref{sec:clustering}, will be valid in different contexts. 

As a matter of fact, some of the internal variations make the perspectives overlap in particular scenarios.
For instance, one can compare all algorithms on simple, undirected LFR benchmark graphs~\cite{Lancichinetti2008}.
However, the LFR benchmark clearly imposes a density-based notion of communities.
Similarly, for simple undirected networks, optimizing Modularity corresponds to the inference of a particular SBM~\cite{Newman2016} or may be reinterpreted as a diffusion process on a graph~\cite{Delvenne2013}.
Nevertheless, this overlap of concepts, typically present on unweighted undirected networks, is only partial, and breaks down, for example, for directed, weighted networks, or for more complex dynamics.

\section{Conclusions}

In summary, no general purpose algorithm will ever serve all applications or data types~\cite{Peel2016}, because each perspective emphasizes a particular core aspect: a cut-based method provides good separation of balanced groups, a clustering method provides strong cohesiveness of groups with high internal density, stochastic block models provide strong similarity of nodes inside a group in terms of their connectivity profiles, and methods that view communities as dynamical building blocks aim to provide node groups that influence or are influenced by some dynamics in the same way.
As more and more diverse types of data are collected, leading to ever more complex network structures, including directed~\cite{Malliaros2013}, temporal~\cite{Holme2012,Sekara2016}, multi-layer or multiplex networks~\cite{Boccaletti2014}, the differences between the perspectives presented here will become even more striking---the same network might have multiple valid partitions depending on what question about the network we are interested in.
We might moreover not only be interested in partitioning the nodes, but also in partitioning edges~\cite{Ahn2010}, or even motifs~\cite{Benson2016}.
Rather than striving to find a `best' community-detection algorithm for a better understanding of complex networks, we argue for a more careful treatment of what network aspects that we seek to understand when applying community detection.

\section*{Acknowledgements}

\noindent We thank Aaron Clauset, Leto Peel and Daniel Larremore for fruitful discussions. 
M.R. was supported by the Swedish Research Council grant 2012-3729.
MTS, JCD, and RL acknowledge support from: FRS-FNRS; the Belgian Network DYSCO (Dynamical Systems, Control and Optimisation) funded by the Interuniversity Attraction Poles Programme initiated by the Belgian State Science Policy Office; and the ARC (Action de Recherche Concerte) on Mining and Optimization of Big Data Models funded by the Wallonia- Brussels Federation.

\bibliography{references}

\end{document}